\documentclass[12pt,preprint]{aastex}

\shorttitle{\ion{Fe}{2} and \ion{Mg}{2} in NLS1s from the SDSS}
\shortauthors{Leighly and Moore}

\received{2004 April 30}
\begin{document}


\title{\ion{Fe}{2} and \ion{Mg}{2} in Luminous, Intermediate-Redshift
  Narrow-line Seyfert 1 Galaxies from the Sloan Digital Sky Survey}


\author{Karen M. Leighly and John R. Moore\altaffilmark{1}}
\affil{Homer L.\ Dodge Department of Physics and Astronomy, The
  University of   Oklahoma, 440 W. Brooks St., Norman, OK 73019}
\email{leighly@nhn.ou.edu}
\altaffiltext{1}{Current Address: Department of Physics \& Astronomy,
  University of Wyoming, Dept.\ 3905, Laramie, WY 62071}



\begin{abstract}
We present results from analysis of spectra from a sample of $\sim
900$ quasars from the Sloan Digital Sky Survey.  These objects were
selected for their intermediate redshift ($1.2<z<1.8$), placing
\ion{Mg}{2} and UV \ion{Fe}{2} in the optical band pass, relatively
narrow \ion{Mg}{2} lines, and moderately good signal-to-noise-ratio
spectra. Using a maximum likelihood analysis, we discovered that there
is a significant dispersion in the \ion{Fe}{2}/\ion{Mg}{2} ratios in
the sample.  Using simulations, we demonstrate that this range, and
corresponding correlation between \ion{Fe}{2} equivalent width and
\ion{Fe}{2}/\ion{Mg}{2} ratio, are primarily a consequence of a larger
dispersion of \ion{Fe}{2} equivalent width (EW) relative to
\ion{Mg}{2} EW.  This larger dispersion in \ion{Fe}{2} EW could be a
consequence of a range in iron abundance, or in a range of \ion{Fe}{2}
excitation.  The latter possibility is supported by evidence that
objects with weak (zero) \ion{C}{2}]~$\lambda 2325$ equivalent width
are likely to have large \ion{Fe}{2}/\ion{Mg}{2} ratios.   We discuss
physical effects  that could produce a range of
\ion{Fe}{2}/\ion{Mg}{2} ratio. 
\end{abstract}




\section{Introduction}

The properties of UV \ion{Fe}{2} and \ion{Mg}{2} in Active Galactic
Nuclei (AGN) are important for several reasons.  Luminous quasars are
probes of the early Universe.  As discussed by \citet{hf93} and
others, the production of iron is thought to lag that of the $\alpha$
elements, including magnesium, due to different formation mechanisms:
magnesium and about half of the iron \citep{nnk92} are produced in
supernovae from massive, rapidly-evolving stars, while the remainder
of the iron is produced largely in Type 1a supernova which involve
accretion onto a white dwarf star, a process which requires
approximately 1 billion years.  Observation of an evolution of [Fe/Mg]
with redshift could constrain the time of the first burst of star
formation in the Universe.  The UV \ion{Fe}{2} and \ion{Mg}{2} line
emission are found conveniently near one another in the rest-UV
bandpass, and their atomic properties are sufficiently similar that
they should be strongly emitted from gas under similar physical
conditions. Therefore it was thought that the \ion{Fe}{2}/\ion{Mg}{2}
ratio could be an abundance diagnostic. No clear evidence for
\ion{Fe}{2}/\ion{Mg}{2} ratio evolution has been observed yet
\citep[e.g.,][]{dietrich03}.  Thus, the first star formation occurred
very early; alternatively, massive amounts of iron may have been
produced in the first very massive stars \citep{hw02}.

The study of \ion{Fe}{2} is also relevant for understanding
Narrow-line Seyfert 1 galaxies (NLS1s).  One of the
criteria used to identify NLS1s, along with their narrow H$\beta$ and
weak forbidden lines, is their frequently-strong optical \ion{Fe}{2}
emission \citep{op85,goodrich89}. NLS1s lie at one
end of the \citet{bg92} Eigenvector 1, and the strength of
optical \ion{Fe}{2} is a primary participant in that
eigenvector.  Also, we are encouraged to use NLS1s to study
\ion{Fe}{2} for pragmatic reasons.  The numerous \ion{Fe}{2} emission
lines form a pseudocontinuum, but since the \ion{Fe}{2} line widths
are correlated with the H$\beta$ widths \citep[e.g.,][]{bg92},
the characteristic shape of the pseudocontinuum and even emission from
individual multiplets can be identified in spectra from objects with
narrow H$\beta$ lines.  In broad-line objects, the \ion{Fe}{2} is smeared,
making it more difficult to study.

Understanding \ion{Fe}{2} could be quite important for understanding
AGN broad-line region (BLR) emission in general.  \ion{Fe}{2}
comprises up to one third of the line emission, so it is an important
coolant \citep{joly93}. The Fe$^+$ ion is sufficiently complicated that it
can potentially be a diagnostic of density, column density,
turbulence, temperature, and continuum shape.  But at the same time,
\ion{Fe}{2} emission is difficult to understand because it is so
complicated, and although sophisticated models are under development
\citep{verner99,sp03,verner03},
none fully account for the required complex atom, ionization balance
and radiative transfer.

The problem of \ion{Fe}{2} emission in AGN spectra has been around
for more than 25 years.  Interest in this complicated problem has
recently increased, stimulated by the availability of high
signal-to-noise ratio IR spectra, the potential for determining the
epoch of the first star formation, and sufficient computing power for
appropriately complex models.  In this paper we present some of the results
of a study of the properties of UV \ion{Fe}{2} and \ion{Mg}{2} in a
large sample of intermediate-redshift narrow-line quasars from the
Sloan Digital Sky Survey (SDSS) Data Release 1
\citep{ab03,richards02}.  Additional details and other results from
an extended sample will be presented in Leighly et al., in preparation.

\section{Data and Reduction}

A strength of the Sloan Digital Sky Survey  for AGN
emission-line studies is that it allows construction of large,
uniformly selected samples.  We initially selected all quasars from
the SDSS Data Release 1 (DR1) that had catalogued redshifts between
1.2 and 1.8, so that \ion{Mg}{2} and UV \ion{Fe}{2} fall squarely in
the SDSS spectra.  We further selected objects that have \ion{Mg}{2}
FWHM $<3000\, \rm km\,s^{-1}$ as measured by the reduction pipeline.
These selection criteria produced a sample of 
more than 1700 objects.  These were examined visually, and objects
that were misclassified as having narrow \ion{Mg}{2}, usually because
of absorption lines, and spectra with low signal-to-noise ratios were
removed. This left 924 spectra for more detailed analysis.  NLS1s are
generally classified by their optical properties; however, it has been
shown that the velocity widths of \ion{Mg}{2} and H$\beta$ are
correlated \citep{mj02}. Therefore, our sample comprises a
large collection of intermediate-redshift luminous NLS1s.
The redshifts of the 924 spectra were refined by cross
correlation with a preliminary composite spectrum of narrow-line
quasars developed from the SDSS Early Data Release spectra.  Then,
following \citet{dietrich02}, we developed a semi-automatic program
to remove the portions of spectra contaminated by absorption lines,
bad pixels, noisy background subtraction, and cosmic rays.   These
points were ignored in further analysis and construction of the
composite spectra.   

A parameter we use later in the analysis is a measurement
characterizing the signal-to-noise ratio in the spectrum.  In each
spectrum, the signal-to-noise ratio is a function of the wavelength,
but we needed a single parameter to characterize the signal-to-noise
ratio in the wavelength range of interest.  We compute the mean
signal-to-noise ratio in 30-point bins between 2200 and 2600 \AA\/, and
use the median of these as the signal-to-noise ratio characterizing
the spectrum. This procedure has the advantages that the wavelength
range chosen includes no strong emission lines, yet is in the region
of interest. It is also robust to bad points in the spectrum, since
they are not present in every 30-point string.

Our aim was to measure the properties of UV \ion{Fe}{2} and
\ion{Mg}{2}.  To measure the flux of the \ion{Fe}{2} pseudocontinuum,
we followed the procedure previously used by a number of authors
\citep[e.g.,][]{bg92,cb96,forster01,leighly99,lm04,dietrich02,dietrich03}.
We first developed a UV \ion{Fe}{2} template from the {\it 
HST} spectrum from the prototypical Narrow-line Quasar I~Zw~1,
following \citet{vw01}: we subtracted a power law
identified at the relatively line-free regions near 2200\AA\/ and
3050\AA\/; absorption lines, and prominent emission lines not
attributable to \ion{Fe}{2} were then subtracted.

A potential problem with this template analysis is that we do not know
the flux of the \ion{Fe}{2} pseudo-continuum lying directly under
\ion{Mg}{2}.  Traditionally, the template has been set equal to zero
in that wavelength range \citep[e.g.,][]{vw01,dietrich03}.  The
possible problem with that procedure is that the gap in the template
is fit as part of the \ion{Mg}{2} line, and this can affect its width
and flux.  Therefore, we develop two templates, one in which the
\ion{Fe}{2} flux is assumed to drop to zero under \ion{Mg}{2}
(referred to as the ``traditional'' template), and another in which
the \ion{Fe}{2} flux is assumed to be approximately the same under
\ion{Mg}{2} as it is adjacent to the line (referred to as the ``new''
template).  Fig.\ 1 shows fits to the composite spectrum with both of
these templates; the lower flux in the \ion{Mg}{2} line for the new
template fit can clearly be seen.  Thus, we perform much of the
analysis in parallel using both templates.  Results that are the same
for both templates should be robust to the systematic uncertainty of
our lack of knowledge of the \ion{Fe}{2} flux under \ion{Mg}{2}, at
least to first order. 

We then used the IRAF spectral fitting program {\it Specfit} to model
the spectra \citep{kriss94} between 2200 and 3050\AA\/.  The model
consisted of a power law, the \ion{Fe}{2} pseudocontinuum, the
\ion{Mg}{2} doublet, \ion{C}{2}]$\lambda 2325$, and
\ion{Fe}{3}$\lambda\lambda 2419.3, 2438.9$.  The \ion{Mg}{2} doublet
components were constrained to have equal flux and width, and fixed
separation.  The \ion{C}{2}] and \ion{Fe}{3} lines were weak, and not
present in all objects; therefore we fixed the wavelengths to their
rest wavelengths, and fixed the widths to $2000\, \rm km \, s^{-1}$,
with the aim of measuring their fluxes and equivalent widths but no
other properties. The {\it Specfit} output yields a measurement and
statistical uncertainty for each parameter.

For a majority of the spectra, the I~Zw~1 template modeled the
iron fairly well. Objects in which the pseudocontinuum shape appeared
significantly different than that of I~Zw~1, and objects with
exceptional \ion{Fe}{2}/\ion{Mg}{2} ratios will be discussed in
Leighly et al.\ in prep.

We measured the luminosity of the continuum at 2500\AA\/ ($H_0=70\rm\,
km\,s^{-1}$, $\Omega_M=0.3$, $\Lambda_0=0.7$), the \ion{Fe}{2} and
\ion{Mg}{2} fluxes, the \ion{Mg}{2} and \ion{Fe}{2} equivalent widths,
and the \ion{Mg}{2} velocity width.  We computed the black hole mass
using the formula presented by \citet{mj02}, based on the
luminosity at 3000\AA\/ and the velocity width of \ion{Mg}{2}.  We
estimate the bolometric luminosity using $\lambda L_\lambda$ at
2500\AA\/ bolometric correction factor of 5.26 \citep[their median
  value]{elvis94}.  We then compute $L/L_{Edd}$.  This will be
proportional to \.M$/M_{BH}$, assuming that the efficiency of
conversion of gravitational potential energy to radiation is the same
in all objects.  We then discarded 21 more objects due to low
signal-to-noise ratio, leaving a sample of 903 objects.

\section{Analysis}

\subsection{Maximum Likelihood Analysis Part 1}

We first determine whether there is significant variance in the
parameters that we measure, or whether the data is consistent with a
constant.  To do this, we use the maximum likelihood method
\citep{mgwzs88} to determine the mean and dispersion and uncertainties
on the following parameters: \ion{Mg}{2} equivalent width (EW),
\ion{Fe}{2} EW, \ion{Fe}{2}/\ion{Mg}{2}, \ion{Mg}{2}/\ion{Fe}{2},
\ion{Mg}{2} FWHM, $L_{2500}$, $M_{BH}$, and $L/L_{Edd}$.  We also
investigate the distributions of the luminosities of \ion{Mg}{2} and
\ion{Fe}{2}, because the equivalent width is a function of two
physical parameters: the line flux in comparison with the
continuum flux, and the covering fraction.  The maximum
likelihood method computes the best estimate of the mean and
dispersion of the data, accounting explicitly for the errors in the
data. Thus, a non-zero dispersion implies real variance in the data,
not just statistical fluctuations.

The luminosities have somewhat of a large and asymmetric spread, and
therefore we discuss the log of these values.    Taking the log of a
value makes the uncertainties nominally non-symmetric; however, we
need a symmetric error for further analysis.  We estimate the errors
in the logarithm of the value using propagation of errors, taking the
first term in the Taylor expansion.  To determine the validity of the
estimation, we compute the ratio of the second term with the first
term.  In all cases, that ratio is less  than 4\%, indicating that the
errors are symmetric to within 4\%.  We deem this acceptably small
uncertainty, and henceforth use the first term in the expansion as a
symmetric error. 

Table 1 lists the results of the maximum likelihood analysis.  All of
the parameters that we are interested in have dispersions
significantly different from zero.  This means that there is a real
range of values of these parameters.  

It is particularly interesting that \ion{Fe}{2}/\ion{Mg}{2} is not
consistent with a constant. Fig.\ 2 shows the histogram of the
\ion{Fe}{2}/\ion{Mg}{2} values, and Fig.\ 3 shows the maximum
likelihood contours.  It is  interesting to note that the
mean value for the traditional template of 3.93 is quite similar to
that found for high redshift quasars \citep{dietrich03}.   All of
these objects 
have redshifts between 1.2 and 1.8, and at this point, evolution of
the iron and magnesium ratios should have ceased \citep{hf93}.
Assuming uniform evolution of elements in the host galaxies, 
and uniform excitation of \ion{Fe}{2} and \ion{Mg}{2} in all objects,
the ratio should be consistent with a constant, in contrast with what
we find.  We therefore next performed several analyses to determine
the origin of the range of values of \ion{Fe}{2}/\ion{Mg}{2}.  

The histogram and contour for the new template are shifted toward
higher \ion{Fe}{2}/\ion{Mg}{2} compared with the results for the
traditional template.  Comparing maximum-likelihood means for the
\ion{Mg}{2} and \ion{Fe}{2} equivalent widths, we find that the
difference lies in the \ion{Mg}{2} equivalent widths; they are
systematically smaller for the new template, while the means of the
\ion{Fe}{2} equivalent widths are consistent between the two
templates.  This is expected, because we expect \ion{Mg}{2} equivalent
width to grow in step with \ion{Fe}{2} equivalent width for the traditional
template.  This systematic propagates to a larger dispersion in
\ion{Fe}{2}/\ion{Mg}{2} ratio for the new template; the dispersion to
mean ratio is 33\% for the new template, and 24\% for the traditional
template.

\subsection{Correlations}

Table 2 presents the Spearman rank correlation coefficient $r$ for all
of the parameters discussed above.  We note several apparently strong
correlations with $|r|>0.5$: between the \ion{Fe}{2} and
\ion{Mg}{2} equivalent widths, between the line luminosities, between
the \ion{Fe}{2} line equivalent widths and their luminosities, between the
\ion{Fe}{2}/\ion{Mg}{2} ratio and the \ion{Fe}{2} equivalent width and
luminosity, between the continuum luminosity and line luminosity, and
between the Eddington ratio, black hole mass, the luminosity, and
\ion{Mg}{2} FWHM.  

Do these correlations give us any physical 
insight, especially for the question of the origin of the range of
\ion{Fe}{2}/\ion{Mg}{2} ratios observed?  Specifically, we expect the
line and continuum luminosities to be 
correlated in a flux-limited sample.  The fluxes are correlated, and
the narrow redshift range does not decorrelate the luminosities.  The
line equivalent widths are correlated. Since the \ion{Fe}{2} and
\ion{Mg}{2} emission should occur in the same gas, this correlation is
expected and may indicate a range of covering fractions; it is also a
function of the correlations in fluxes.  In addition, the Eddington
ratio parameter and black hole mass should be correlated with the
continuum luminosity and the \ion{Mg}{2} FWHM because they are
functions of these two parameters. 

There are two potentially physically interesting correlations.  The
first is an 
anticorrelation between the \ion{Mg}{2} equivalent width and the
Eddington ratio.  The Eddington ratio parameter is a
function of the continuum luminosity and the \ion{Mg}{2} FWHM, so one
may think that this correlation is a consequence of an anticorrelation
between the continuum luminosity and the line equivalent width \citep[the
Baldwin effect;][]{bal77} and the correlation between the equivalent width
and the velocity width that has been seen previously
\citep[e.g.,][]{gaskell95}.  However, neither of these latter two
correlations are particularly strong in this sample, nor are they as
strong as the equivalent width and Eddington ratio correlation.   

Also interesting is the correlation between the
\ion{Fe}{2}/\ion{Mg}{2} ratio and the \ion{Fe}{2} equivalent width
($r=0.62$ for both the traditional and new templates). At first
glance, this may appear to be a trivial correlation, because both
parameters are positively correlated with the \ion{Fe}{2} 
flux.  However, there is no corresponding
correlation ($r=0.13$ and 0.33 for the traditional and new templates,
respectively) between \ion{Mg}{2} EW and \ion{Mg}{2}/\ion{Fe}{2}. 

It is still possible that this correlation is spurious.  It is
expected that the equivalent width of \ion{Mg}{2} should be relatively
reliably measured, statistically speaking, because it is a sharp
feature.  The uncertainty in the equivalent width of \ion{Fe}{2} 
may be larger because it is a broad feature and competes with the
continuum.  Indeed, the mean relative error of the  \ion{Fe}{2} equivalent
widths is larger than that of the \ion{Mg}{2} equivalent widths (6.4\%
vs 4.0\% for the traditional template; 6.4\% vs 4.4\% for the new template).
So, if the \ion{Fe}{2}/\ion{Mg}{2} ratio were intrinsically constant,
a positive fluctuation in \ion{Fe}{2} for a particular object would
give a positive fluctuation both the ratio and the \ion{Fe}{2}
equivalent width, with a similar result for a negative fluctuation.
The result would be a positive correlation between these parameters, as
is seen.   An accompanying fluctuation in \ion{Mg}{2} would not give
rise to as strong of a correlation because it is more securely measured.

On the other hand, a positive correlation might be observed if its
origin is physical.  Specifically, if the \ion{Fe}{2} equivalent width
varies more in the sample than the \ion{Mg}{2} equivalent width, as a
result of a differences in \ion{Fe}{2} excitation or iron abundance, a
positive correlation between \ion{Fe}{2}/\ion{Mg}{2} ratio and
\ion{Fe}{2} equivalent width would also be seen.  Indeed, from the
maximum likelihood analysis, we find that the dispersion relative to
the mean is larger for the \ion{Fe}{2} equivalent width compared with
the \ion{Mg}{2} equivalent width (35\% vs 26\% for the traditional 
template; 35\% vs 27\% for the new template).  Since the maximum
likelihood analysis accounts for the measurement error, these numbers
should reflect real differences in the dispersions of these
parameters.

In the next two sections we use simulations and additional maximum
likelihood analysis to try to determine whether or not the range of
\ion{Fe}{2}/\ion{Mg}{2} originates in statistical error, or arises in
a larger range in values of \ion{Fe}{2} EW relative to \ion{Mg}{2} EW
that may reflect differences in abundance or excitation.

\subsection{Simulations}

The maximum likelihood analysis indicates that the
\ion{Fe}{2}/\ion{Mg}{2} ratio varies significantly in the sample, and
the larger dispersion of \ion{Fe}{2} equivalent width relative to
\ion{Mg}{2} equivalent width suggests that this is the cause of
the dispersion in the ratio.  However, the uncertainties in the
measurements of \ion{Fe}{2} EW are about 1.6 times larger than those in
\ion{Mg}{2} EW, a fact that can also cause a correlation.  In this
section we present simulations that are designed to test the influence
of each factor in turn.   We only describe simulations on the traditional 
template data; the results for the new template are essentially the
same. 

First we construct samples of simulated \ion{Fe}{2} and \ion{Mg}{2}
equivalent widths and corresponding \ion{Fe}{2}/\ion{Mg}{2} ratios
that have essentially the same distributions as the observed data.
The simulated data are constructed in three steps.  First, we
construct Gaussian distributions for the \ion{Fe}{2} and \ion{Mg}{2}
equivalent widths that have means equal to the observed means, and
variances equal to the observed variances times scale factors that are
determined as described below. 

The next step is construction of the errors.  The errors are not
simply related to the data because we are using the derived quantity
equivalent width; furthermore, the uncertainties in the
flux are also not simply related to the flux, because observations are
of different length and are made under differing conditions. To zeroth
order, the errors are correlated with the data.  We assume a linear
relationship between the log of each parameter and the log of its
error, fit for the slope of the distribution, $\beta$, and use the result
to construct the initial errors on the simulated data.  Then, because
the errors themselves have a large spread for a given value of the
data, we randomize them further by adding a Gaussian random variable
that has magnitude equal to a constant factor, determined empirically,
times the data.  The result is that the constructed errors have a
variance in $\log_{10}(error)-\beta\log_{10}(data)$ that is only
slightly lower  than observed. 

We next add noise to the data.  The noise amplitude is set equal to
the square root of the mean square uncertainties.  

At this point in the simulation, we have a large number of pairs of
\ion{Fe}{2} and \ion{Mg}{2} equivalent widths or luminosities.
However, not all pairs are valid, because the ratio of \ion{Fe}{2} to
\ion{Mg}{2} is not arbitrary in the observed data.  We assume a
Gaussian distribution of \ion{Fe}{2}/\ion{Mg}{2} ratio that has the
mean and standard deviation of the real data.   We partition this
distribution into bins of width 0.1 and discretize it.  We then pack
the distribution with pairs of \ion{Fe}{2} and \ion{Mg}{2} equivalent
width or luminosity, rejecting pairs that are outside the distribution
or that fall in a particular bin that is already filled.

Forcing the simulated \ion{Fe}{2}/\ion{Mg}{2} distribution to match
the observed \ion{Fe}{2}/\ion{Mg}{2} distribution narrows the
distributions of the simulated \ion{Fe}{2} and \ion{Mg}{2} equivalent
widths.  To account for this, we broaden the initial distribution by
the scale factor mentioned above.  To determine the value of the scale
factors, which are different for \ion{Mg}{2} EW and \ion{Fe}{2} EW, we
require that the variance in the simulated data match that of the
original data.  These scale factors are determined by running the 
distribution simulation program 100 times each for a range of initial
scale factors, and then determining which scale factor produces the
observed variance on average.

Various statistics for the real and these simulated (Simulation 1)
data are given in Table 3.  We see that, as intended, the
distributions, specifically the mean error/data and the maximum
likelihood dispersion/mean, are essentially identical. 

We list in Table 3 the Spearman rank correlation coefficients between
the \ion{Fe}{2} EW and the \ion{Fe}{2}/\ion{Mg}{2} ratio, and between
the \ion{Mg}{2} EW and the \ion{Mg}{2}/\ion{Fe}{2} ratio.  For both
the real data and this first set of simulated data, the correlation
coefficient between the \ion{Fe}{2} EW and the \ion{Fe}{2}/\ion{Mg}{2}
ratio is much larger than the correlation coefficient between the
\ion{Mg}{2} EW and the \ion{Mg}{2}/\ion{Fe}{2} ratio, as expected,
because both the dispersion in \ion{Fe}{2} equivalent widths and the
mean relative error in \ion{Fe}{2} EW are larger than those from
\ion{Mg}{2}. 

The second set of simulations (Simulation 2) adjusts the dispersion of
the \ion{Fe}{2} and \ion{Mg}{2} equivalent widths so that ratio of the
maximum likelihood values of the dispersion and the mean are
approximately equal.  This is done 
by increasing the width of the \ion{Mg}{2} EW dispersion, and
decreasing the width of the \ion{Fe}{2} EW dispersion.  The relative
errors are kept the same as the original data; now this is the only
difference between the \ion{Mg}{2} and \ion{Fe}{2} simulated
equivalent widths.  We see that the correlation coefficient between
\ion{Fe}{2} EW and \ion{Fe}{2}/\ion{Mg}{2} ratio drops, and the
correlation coefficient between \ion{Mg}{2} EW and the
\ion{Mg}{2}/\ion{Fe}{2} ratio increases compared with those from
Simulation 1 by a large amount, and now they are relatively close
together.  This suggests that the distribution of the equivalent
widths influences the correlations strongly.

In the third set of simulations (Simulation 3), we increase the
\ion{Mg}{2} equivalent width uncertainty by a factor of 1.6, so that
the mean of the relative error is the same for both the \ion{Mg}{2}
and \ion{Fe}{2} equivalent widths.  But we leave the maximum
likelihood dispersion/mean of the equivalent widths the same as the
real data.  In this case, we find  very little difference in
the correlation coefficients compared with Simulation 1, suggesting
that the different relative errors in the data influences the
correlations very little.

In the final set of simulations (Simulation 4), we make both the
dispersions of the \ion{Fe}{2} and \ion{Mg}{2} equivalent widths, and
the relative uncertainty on the \ion{Fe}{2} and \ion{Mg}{2} equivalent
widths equal.  In this case we match the straight standard deviation
divided by the mean, rather than the maximum likelihood value because
although the mean relative uncertainties are set equal, the
distribution is somewhat different (see above), and that enters into
the maximum likelihood estimation.  Regardless, the maximum likelihood
estimates of the dispersion/mean are very close (0.29 vs 0.30 for the
\ion{Mg}{2} EW and \ion{Fe}{2} EW, respectively).  We find that the
correlation coefficient between \ion{Fe}{2} EW and
\ion{Fe}{2}/\ion{Mg}{2} is the same as \ion{Mg}{2} EW and
\ion{Mg}{2}/\ion{Fe}{2}, as expected, since the factors that influence
the correlation coefficient are now equal.

Our conclusions from these simulations is that while both the relative
uncertainty and the dispersion of the data can result in positive
correlations, the influence of the distribution is much more important
in these data; the relative uncertainty has very little influence.  So
we conclude that the reason that the correlation between  \ion{Fe}{2} EW
and \ion{Fe}{2}/\ion{Mg}{2} is stronger than that between \ion{Mg}{2}
EW and \ion{Mg}{2}/\ion{Fe}{2} is because the dispersion in \ion{Fe}{2} EW is
larger than that of \ion{Mg}{2} EW, and {\it not} because the relative
uncertainties in the data are larger for \ion{Fe}{2} EW compared with
\ion{Mg}{2} EW.

\subsection{Maximum Likelihood Analysis Part 2}

The maximum likelihood analysis over the whole sample given in Table 1
shows that  the dispersion is significant in \ion{Mg}{2} and
\ion{Fe}{2} equivalent widths, as well as in \ion{Fe}{2}/\ion{Mg}{2}
ratios.  In principle, the dispersion in \ion{Fe}{2}/\ion{Mg}{2} could
be produced by a larger spread in \ion{Mg}{2}, or a larger spread in
\ion{Fe}{2}.  The fact that the dispersion relative to the mean is
larger in \ion{Fe}{2} compared with \ion{Mg}{2} (35\% compared with 26\%
for the traditional template, 35\% compared with 27\% for the new
template) suggests that \ion{Fe}{2} is the culprit.  In this section,
we investigate this further using a maximum likelihood analysis.

We sort the spectra according to \ion{Mg}{2} equivalent width, and
divide into nine bins, according to the value of that parameter.  We
then compute the maximum likelihood estimate of the
\ion{Fe}{2}/\ion{Mg}{2} ratio in each bin.  We do the same for
\ion{Fe}{2} equivalent widths.  

The results for the traditional and new templates are shown in Fig.\
4.  They show that the mean and dispersion of \ion{Fe}{2}/\ion{Mg}{2}
has much different behavior depending on whether they are computed
from spectra in bins chosen by their \ion{Mg}{2} equivalent width or
their \ion{Fe}{2} equivalent width.  Bins chosen by \ion{Fe}{2}
equivalent width yield a broad range of maximum-likelihood mean
\ion{Fe}{2}/\ion{Mg}{2} ratios.  Bins chosen by \ion{Mg}{2} equivalent
width yield maximum-likelihood mean \ion{Fe}{2}/\ion{Mg}{2} ratios
almost independent of the \ion{Mg}{2} equivalent width.  Furthermore,
the dispersion of  \ion{Fe}{2}/\ion{Mg}{2} for bins chosen by
\ion{Mg}{2} equivalent width is similar to the dispersion of
\ion{Fe}{2}/\ion{Mg}{2} in the whole sample. This means that in each
\ion{Mg}{2} EW bin, a broad range of \ion{Fe}{2}/\ion{Mg}{2} is
present, similar to the whole sample.  In contrast, at least for
intermediate values of \ion{Fe}{2} equivalent width, the dispersion in
the \ion{Fe}{2}/\ion{Mg}{2} ratio drops compared with the whole
sample; objects in these bins are more likely to have the same value
of \ion{Fe}{2}/\ion{Mg}{2} ratio.  These results imply that
\ion{Fe}{2} influences the \ion{Fe}{2}/\ion{Mg}{2} ratio more than
\ion{Mg}{2}.  Recall that the maximum likelihood analysis accounts for
the measurement errors in the equivalent widths and ratios, so these
results cannot be attributed to the slightly larger relative
uncertainties on \ion{Fe}{2} parameters. 

\subsection{Composite Spectra}

We next study composite spectra to see if we can obtain some insight
on the physical origin of the range of \ion{Fe}{2}/\ion{Mg}{2} ratios.  

First, we sort the observed \ion{Fe}{2}/\ion{Mg}{2} values, and divide
them into nine bins of 100 objects.  We then construct composite spectra of
the objects in each bin with signal-to-noise ratio greater than the
sample median.  The resulting composite spectra are composed of
between 40 and 65 individual spectra.   

We then fit each spectrum using the same model as before, except
we add an additional line with width fixed at $2000 \rm \, km\,s^{-1}$
at 2745.72\AA\/.  This component fits the contribution of \ion{Fe}{2}
UV 62,63 that sometimes appears as a spike in the spectrum.  We fit
the model twice, using both the traditional and new templates.   The
fit parameters are plotted as a function of measured \ion{Fe}{2}/\ion{Mg}{2}
ratio in Fig.\ 5.  The fits for the spectra from the lowest and
highest \ion{Fe}{2}/\ion{Mg}{2} ratio bins are shown in Fig.\ 6.

The results are illuminating.  First of all, we confirm our suspicions
about the influence of the template on the \ion{Mg}{2} properties.
While the \ion{Mg}{2} equivalent width drops with increasing
\ion{Fe}{2}/\ion{Mg}{2} ratio for both templates, the decrease is less
for the traditional template: the standard deviations divided by the
means of the composite-spectra \ion{Mg}{2} EW are 0.047 and 0.11 for
the traditional and new templates, respectively.  However, the
increase in the \ion{Fe}{2} equivalent width as
\ion{Fe}{2}/\ion{Mg}{2} increases is larger than the decrease in
\ion{Mg}{2} equivalent width: the standard deviations divided by the
means are 0.21 and 0.22 for the traditional and new templates
respectively.  This provides additional evidence that the range of
\ion{Fe}{2}/\ion{Mg}{2} observed is primarily a result of a larger
dispersion in \ion{Fe}{2} compared with \ion{Mg}{2}.

There are several other interesting results.  The \ion{Mg}{2} FWHM
decreases with increasing \ion{Fe}{2}/\ion{Mg}{2} by $\sim 500\rm \,
km\, s^{-1}$ and $\sim 700\rm \, km\, s^{-1}$ for the traditional and
new templates, respectively.  The difference between the \ion{Mg}{2}
velocity widths for the two templates is most pronounced for the
largest-width \ion{Mg}{2} lines, as anticipated, but it was less than
$200 \rm\, km\,s^{-1}$ in any case, and is zero for the
narrowest-width lines. This result appears to imply that, like optical
\ion{Fe}{2} \citep[e.g.,][]{bg92}, UV \ion{Fe}{2} is stronger in
objects with narrower lines. 

The smaller lines, \ion{C}{2}], \ion{Fe}{3}, and \ion{Fe}{2} UV 62, 63,
show interesting trends with \ion{Fe}{2}/\ion{Mg}{2} ratio.
\ion{C}{2}] is strongest when the ratio is low, and significantly
weaker when the ratio is high, especially for the two spectra with
largest \ion{Fe}{2}/\ion{Mg}{2}.  This is interesting, because the
properties of this semiforbidden line may give us some clues about
the physical (versus phenomenological) origin of the range in
\ion{Fe}{2}/\ion{Mg}{2} ratio (see \S 4.2).  \ion{Fe}{3} is weaker
when \ion{Fe}{2}/\ion{Mg}{2}  is small, and nearly constant with
higher values of this ratio.  Finally, \ion{Fe}{2} UV 62, 63 is
nearly constant with \ion{Fe}{2}/\ion{Mg}{2}, except it is
significantly stronger for the spectrum with the largest
\ion{Fe}{2}/\ion{Mg}{2} ratio.  

Since \ion{C}{2}] is potentially important for understanding the
origin of the  \ion{Fe}{2}/\ion{Mg}{2} dispersion, we investigate
its properties in another way.  We have fitted this line in all of the
spectra.  It is a weak line, so we can't simply take the measured
equivalent widths at face value; we need to be certain that we are
analyzing meaningful detections.  To find objects with large
\ion{C}{2}], we isolate objects in which  the \ion{C}{2}] flux is not
equal to zero and in which the measured  flux is more than three times
the uncertainty.  This leaves 352 objects.  From these, we make a composite
spectrum from the spectra of the 70 objects with the highest
\ion{C}{2}] equivalent widths; that composite spectrum is shown in the
  left panel of Fig.\ 7.  We also need to compile a sample with low
  values of \ion{C}{2}] equivalent width.  To do this, we use 
the 67 objects that have fitted \ion{C}{2}] equivalent width equal
to zero, and have a signal-to-noise ratio of the spectrum greater than
the sample median.  That spectrum is shown in the right panel of Fig.\ 7.  We
also show the distribution of the high and zero \ion{C}{2}] equivalent
width objects on the \ion{Fe}{2}/\ion{Mg}{2} ratio vs Eddington
ratio plane, in Fig.\ 8.

These analyses confirm our finding from the
\ion{Fe}{2}/\ion{Mg}{2}-sliced composite spectra: on average, low 
(zero) \ion{C}{2}] EW objects have high \ion{Fe}{2}/\ion{Mg}{2}
ratios, and  high \ion{C}{2}] EW objects have low (actually average)
\ion{Fe}{2}/\ion{Mg}{2} ratios.  The measured \ion{Fe}{2}/\ion{Mg}{2}
ratio (traditional template) for the high and zero \ion{C}{2}] EW
composites are 3.5 and 5.1 respectively.  Furthermore, we find a
distinct separation of the locations of these types of objects on the
\ion{Fe}{2}/\ion{Mg}{2} ratio vs.\ Eddington ratio parameter plane.
Objects with high \ion{C}{2}] equivalent widths have average
\ion{Fe}{2}/\ion{Mg}{2} ratios, and objects with high
\ion{Fe}{2}/\ion{Mg}{2} ratios are very likely to  have \ion{C}{2}]
equivalent widths equal to zero. We tentatively interpret this as 
evidence that the \ion{C}{2}] equivalent width and the
\ion{Fe}{2}/\ion{Mg}{2} ratio are coupled such that physical
conditions that cause \ion{C}{2}]   equivalent width to be zero also
  cause the \ion{Fe}{2}/\ion{Mg}{2}   ratio to be high.  Plausible
  candidates for such physical conditions   are discussed below.

\section{Discussion}

We report the results of analysis of the rest frame 2200--3050\AA\/
region in a sample of 903 objects with relatively narrow \ion{Mg}{2}
lines drawn from the Sloan Digital Sky Survey.  We fit the spectra
with a model consisting of seven components: a powerlaw continuum, the
iron template, and six Gaussians fit to \ion{Mg}{2}~$\lambda\lambda
2796.4,2803.5$, \ion{C}{2}]~$\lambda 2325.0$, and
  \ion{Fe}{2}~$\lambda\lambda  2419.3, 2438.9$.   We used two
  \ion{Fe}{2} templates that treat the region under \ion{Mg}{2}
  differently in order to test the effect of our lack of knowledge of
  the \ion{Fe}{2} flux in that region on the results.  We discovered
  that there is a significant range of \ion{Fe}{2}/\ion{Mg}{2} ratios
  in the sample: the maximum likelihood means and 1-sigma dispersions
  are $3.93 \pm 0.95$ (traditional \ion{Fe}{2} template) and $5.28 \pm
  1.72$ (new \ion{Fe}{2} template).  Through simulations, we show that
  this range, and corresponding correlation between \ion{Fe}{2}
  equivalent width and \ion{Fe}{2}/\ion{Mg}{2} ratio, are primarily a
  consequence of a larger dispersion of \ion{Fe}{2} EW relative to
  \ion{Mg}{2} EW.  This larger dispersion in \ion{Fe}{2} EW could be a
  consequence of a difference in iron abundance, or it could be a
  consequence of a difference in iron excitation.  The latter
  possibility is supported by our discovery of a coupling between the
  \ion{C}{2}] equivalent width and the \ion{Fe}{2}/\ion{Mg}{2} ratio. 
Below, we discuss these two possibilities in turn.

\subsection{Iron Abundance}

Could the larger range in \ion{Fe}{2} equivalent widths  originate in
real variation in the relative iron and magnesium  abundances? At the
intermediate redshift range that we investigate, the Universe is
already 3.5--5 Gyr old. The models calculated by \citet{hf93} show
that the abundances of Fe and Mg have almost stopped evolving at this
point. 

\citet{hf92,hf93,hf99} make the case for normal evolution of stellar
populations in galactic nuclei for the element enrichment in QSOs.  If
that is true, we can estimate the spread in [$\alpha$/Fe] expected in
the sample, because the black hole mass is related to the stellar
velocity dispersion, which is related in turn to [$\alpha$/Fe]; this
is related to the mass-metallicity relationship for elliptical
galaxies. 

We first note that the black holes in our sample of relatively
luminous objects are sufficiently large to have  elliptical
hosts.  The mean and dispersion of the black hole masses are $7.3 \pm
2.7 \times 10^7\, M_\odot$ and $7.1 \pm 3.2 \times 10^7\, M_\odot$ for
the traditional and new templates respectively.  The spheroid mass is
related to the black hole mass by $M_{BH}=0.0012 M_{sph}$
\citep{dunlop04}.  Thus, the spheroid masses are expected to be around
$6 \times 10^{10} \, M_\odot$, the size of a typical elliptical
galaxy.

We have estimated the black hole masses for the sample, using the
\citet{mj02} formalism. From these, we estimate the stellar velocity
dispersion for each object using the formula from \citet{tremaine02}:
$\log(M_{BH}/M_\odot)=\alpha+\beta \log(\sigma/\sigma_0)$, where
$\sigma_0=200 \rm\, km\,s^{-1}$, $\alpha=8.13 \pm 0.06$, and
$\beta=4.02 \pm 0.32$.  At this point, uncertainties are simply the
statistical uncertainties propagated through the equations.  We then
find the maximum likelihood mean and dispersion of $\sigma$ to be $169
\pm 13$ and $166 \pm 17 \rm \, km\, s^{-1}$ for the traditional and
new templates respectively.  The small dispersion in these values is a
consequence of the relatively narrow range in black hole masses
inferred in the sample.  We see from Fig.\ 1 in \citet{tmb02} that
this velocity dispersion corresponds to [$\alpha$/Fe] of approximately
$0.18 \pm 0.05$, where the values given are inferred by eye from the
figure. This implies that the estimated disperson/mean in the Mg/Fe
ratio is reasonably expected to be about 17\%.  This is somewhat
smaller than the 1-sigma dispersion/mean estimate for the sample
\ion{Fe}{2}/\ion{Mg}{2} (24\% and 33\% for the traditional and new
templates, respectively).

Our estimate of the range in $\sigma$ expected in the sample accounted
for only statistical uncertainties.  \citet{mj02} estimate that their
black hole masses derived from \ion{Mg}{2} FWHM and $L_{3000}$ is good
to a factor of 2.5.  Our sample may be more uniform than theirs, in
which case this systematic error may be an overestimate.  Regardless,
for the average value of black hole mass of $7.2 \times 10^7 \,\rm
M_\odot$, a factor of 2.5 higher and lower black hole mass would lead
to a $\sigma$ range of 140--210$\, \rm km\, s^{-1}$, corresponding to
only a slightly larger disperson in [$\alpha$/Fe] of approximately
$0.20 \pm 0.1$.  This corresponds to an estimated expected
mean/dispersion in the Mg/Fe ratio of 23\%, comparable to the observed
1-sigma dispersion/mean estimate for the sample
\ion{Fe}{2}/\ion{Mg}{2}.

It is important to realize, however, that a high iron abundance may
not be directly observable in the \ion{Fe}{2}/\ion{Mg}{2} ratio.  In
other words, a factor of three higher iron/magnesium abundance ratio
may not be manifested in a factor of three larger
\ion{Fe}{2}/\ion{Mg}{2}.  This is because in classical photoionization
models, \ion{Fe}{2} has a thermostatic effect
\citep[e.g.,][]{collin00}.  This means that UV \ion{Fe}{2} should not
be very sensitive to abundance differences \citep{verner00,verner03}.
On the other hand, \citet{verner00} shows that not all \ion{Fe}{2}
lines are affected the same way by changes in abundance because of the
accompanying change in optical depth which affects more strongly
nearly-saturated emission lines.  This implies that we might expect
the iron pseudocontinuum to look different in objects with different
[Fe/Mg].

In addition, it has also been suggested that objects with high Eddington
ratio may have high abundances because rapid star formation may
accompany fast growth of the black hole \citep{mathur00}.  This has been
primarily discussed in the context of nitrogen abundances, which,
depending on the star formation model, increase strongly much before
or concurrent with iron \citep{hf93}. Assuming that the
iron originates in primarily in Type 1a supernovae, it should depend
on how long the high accretion rate period has been in progress
because of the $\sim 1\,\rm Gyr$ delay.  In
our sample, $L/L_{Edd}$ parameter is correlated with
\ion{Fe}{2}/\ion{Mg}{2} but not strongly (Fig.\ 8).

In summary, we find that the expected range in  [$\alpha$/Fe] in the
elliptical hosts may be sufficient to explain the 1-$\sigma$
dispersion in the observed \ion{Fe}{2}/\ion{Mg}{2} ratios.  This
assumes that the abundance ratio is linearly manifested in the line
ratio; it may not be, due to the thermostatic effect of \ion{Fe}{2}.
Regardless, the expected range in abundances is {\it not} sufficient to
explain the long tail of objects with very high
\ion{Fe}{2}/\ion{Mg}{2} ratios that can be seen in Fig.\ 2 or Fig.\ 8.

\subsection{\ion{Fe}{2} Excitation}

Analysis of composite spectra compiled from spectra selected by their
\ion{Fe}{2}/\ion{Mg}{2} ratios (\S 3.5) shows that high 
\ion{Fe}{2}/\ion{Mg}{2} ratios are associated with weak \ion{C}{2}]
and strong \ion{Fe}{2} UV 62,63.  Furthermore, we find that objects
with measured \ion{C}{2}] equivalent widths equal to zero are likely
to have enhanced \ion{Fe}{2}/\ion{Mg}{2} ratios.  These results
suggest that the range in \ion{Fe}{2}/\ion{Mg}{2} is a consequence
of different excitation of \ion{Fe}{2} rather than abundances.

What processes might be responsible for the different excitation?  
\citet{vp04} recently show that a high
\ion{Fe}{2}/\ion{Mg}{2} ratio is predicted by their model when the
photoionizing flux and density are high.  They also 
find that there is a tendency for higher-luminosity objects to have
higher \ion{Fe}{2}/\ion{Mg}{2} ratios.  The nature of the link they
infer between the luminosity of an object and the photoionizing flux
at the BLR and its density is unclear, however. For constant
Eddington ratio, the luminosity, black hole mass, and size of the
emission region should all scale together.  For variable Eddington
ratio, these scalings could change, because the accretion geometry
plausibly changes; however, for our sample, we find that the
\ion{Fe}{2}/\ion{Mg}{2} ratio is not strongly correlated with the
Eddington ratio.  In either case, the spectral energy distribution
should change.  Also, \citet{vp04} confine their
discussion to the \ion{Fe}{2}/\ion{Mg}{2} ratio; it is not clear
whether their model could naturally explain the correlation of that
parameter with \ion{Fe}{2} equivalent width that we see.

Alternatively, the difference in the \ion{C}{2}] equivalent widths in the
high and low \ion{Fe}{2}/\ion{Mg}{2} composite spectra, and the
differences in the distributions of objects with high and zero
\ion{C}{2}] equivalent widths (\S 3.5)  suggests differences in optical
depth or density of the emission-line region.  \citet{kk81}, in an
early BLR photoionization model, used the \ion{C}{2}]$\lambda 2326$
line as a column density diagnostic, because this low-ionization line
is emitted deep in the partially-ionized zone, and because, being a
semiforbidden line, it should be less affected by radiative transfer.
Following this argument, it could be concluded that the low
\ion{Fe}{2}/\ion{Mg}{2} objects have a high column density which both
increases the \ion{C}{2}] EW, and decreases the UV \ion{Fe}{2} EW as
part of the UV \ion{Fe}{2} is converted into optical \ion{Fe}{2}.
This view is supported by the anticorrelation of optical and UV
\ion{Fe}{2} observed in the PG quasars \citep{shang03} and we note
that we can see that \ion{C}{2}]$\lambda 2326$ is anticorrelated with
UV \ion{Fe}{2} in their spectral principal component called (SPC)3.

However, this argument may not be complete.  \citet{fp89} show that
\ion{C}{2}] will be optically thick in gas that produces \ion{Fe}{2}.
More important is the fact that  \ion{C}{2}]$\lambda 2326$ has a
rather low critical density of   $3.16  \times 10^{9}\rm\, cm^{-3}$
\citep{hamann02}.  This   suggests density plays a role, with the low
and high   \ion{Fe}{2}/\ion{Mg}{2}  objects   being characterized by
low and high densities, respectively.  High   density increases
\ion{Fe}{2} emission because Fe$^+$ is primarily   excited by
collisions. \ion{Mg}{2} can decrease at high densities as the line
becomes thermalized. Individual \ion{Fe}{2} transitions can be
saturated at high density, but the large number of transitions
available prevents UV \ion{Fe}{2} emission as a whole from becoming
thermalized, so the \ion{Fe}{2} pseudocontinuum increases at high
densities  \citep{verner00}. 

The enhancement of \ion{Fe}{2} UV 62, 63, which are transitions that
have low-lying upper levels, in high \ion{Fe}{2}/\ion{Mg}{2} objects may
imply that differences in excitation may contribute to the
range of \ion{Fe}{2} equivalent widths.  These Fe$^+$ lines have upper
levels near 5.5~eV that could be  excited by additional heating, perhaps
by a mechanical source \citep[e.g.,][]{collin00}.  The photoionizing
flux can influence the emission from the lowest levels, because if the
flux is low, the partially ionized zone is thin, and the lines from
the low-lying levels are not saturated and appear stronger
\citep{verner00}. Microturbulence could also enhance the emission from
the low-lying levels, because it reduces the effective optical depth
\citep{sp03}. Evidence for microturbulence in NLS1s has already 
been seen; the enhancement of \ion{Fe}{3}$\lambda 1914$ in several
NLS1s quite likely arises from pumping by Ly$\alpha$
\citep{lm04,johansson00}.  The upper level energy 
for \ion{Fe}{3}$\lambda 1914$ corresponds to 1214.6\AA\/, which is
1.1\AA\/ from Ly$\alpha$, implying a velocity difference of $274\,\rm
km\,s^{-1}$.    

It is also possible that the spectral energy distribution (SED) can
affect the \ion{Fe}{2} emission.  A SED with strong hard X-ray
emission can increase the depth of the partially-ionized zone, and
increase the \ion{Fe}{2} emission \citep[e.g.,][]{collin00}.  A very
soft SED may also be able to influence the production of \ion{Fe}{2}
emission as well.  A soft SED may increase the depth of the partially
ionized zone by free-free absorption if it is strong in the IR
\citep{fp89}. In a gas photoionized by a very soft SED,
there will be few highly-ionized ions, because the SED lacks the
photons required by their high ionization potential.  The gas will be
dominated by low-ionization ions, such as H$^+$ and Fe$^+$, and Ly$\alpha$
emission will be strong.   If microturbulence is present,
Ly$\alpha$ pumping of \ion{Fe}{2} may be very strong.
\ion{Mg}{2} would not be similarly enhanced. In this case, the
\ion{Fe}{2} pseudocontinuum may be dominated by high-excitation
transitions.  This may be responsible for the ultra-strong \ion{Fe}{2} 
emission observed in the narrow-line quasar PHL~1811 \citep[Leighly et
  al., in prep.]{lhj04} and in some of the SDSS  
objects (Leighly et al., in prep.).  

%% In this section, we use  the \subsection command to set off
%% a subsection.  \footnote is used to insert a footnote to the text.


%% This section also includes several of the displayed math environments
%% mentioned in the Author Guide.

\acknowledgments
We thank Eddie Baron, Darrin Casebeer, Martin Gaskell, Dirk Grupe and
Steve Kraemer for helpful comments, and Mike Loewenstein for helpful
discussions on metallicity in elliptical galaxies. We also thank the
anonymous referee for comments that led to a substantial increase in
rigor in the analysis, and Randi Worhatch for a careful reading.  KML and
JRM gratefully acknowledge support by NASA grant NAG5-10171
(LTSA). Funding for the creation and distribution of the SDSS Archive
has been provided by the Alfred P. Sloan Foundation, the Participating
Institutions, the National Aeronautics and Space Administration, the
National Science Foundation, the U.S. Department of Energy, the
Japanese Monbukagakusho, and the Max Planck Society.  The SDSS Web
site is http://www.sdss.org/.  The SDSS is managed by the
Astrophysical Research Consortium (ARC) for the Participating
Institutions.  The Participating Institutions are The University of
Chicago, Fermilab, the Institute for Advanced Study, the Japan
Participation Group, The Johns Hopkins University, Los Alamos National
Laboratory, the Max-Planck-Institute for Astronomy (MPIA), the
Max-Planck-Institute for Astrophysics (MPA), New Mexico State
University, University of Pittsburgh, Princeton University, the United
States Naval Observatory, and the University of Washington.

\clearpage


\begin{figure}
\plotone{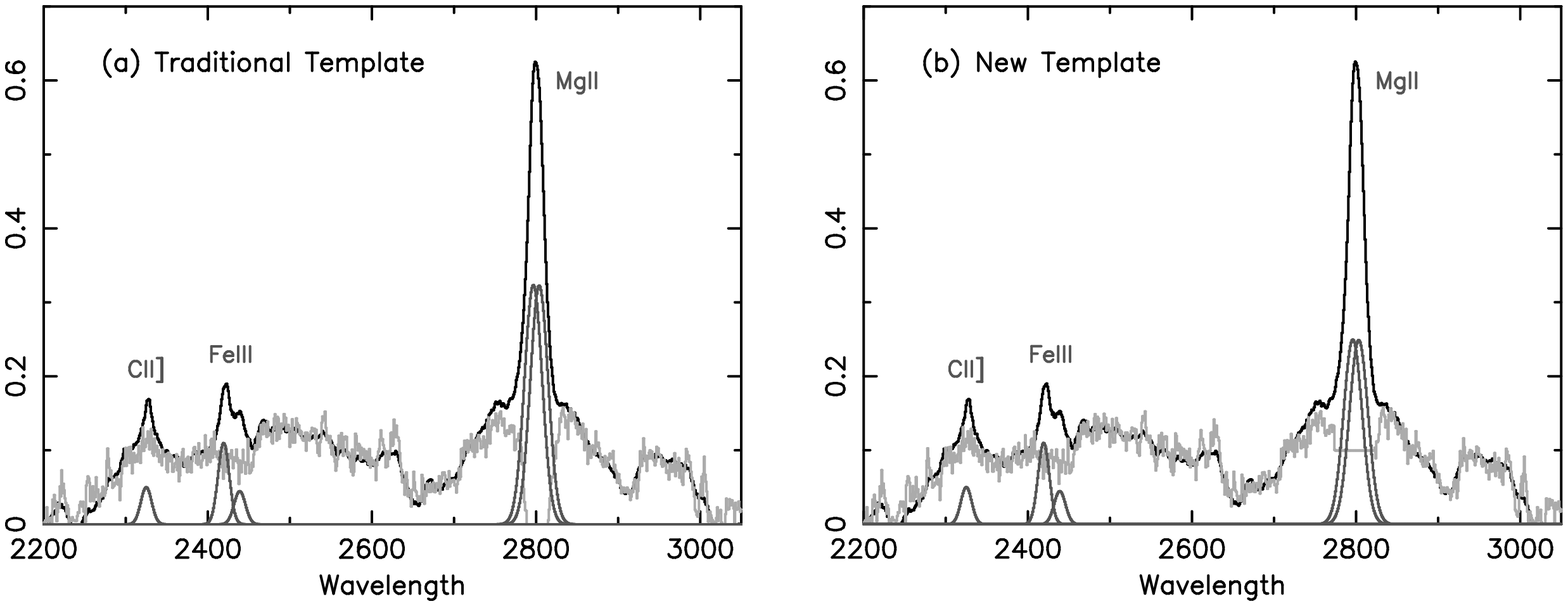}
\caption{Fits to the composite spectrum composed of  spectra with
  signal-to-noise ratio greater than the median.  (a.) The fit using
  the ``traditional'' \ion{Fe}{2} template, in which the \ion{Fe}{2}
  flux under \ion{Mg}{2} is assumed to be zero. (b.) The fit using the
  ``new'' template, in which the \ion{Fe}{2} flux under \ion{Mg}{2} is
  assumed to be comparable to that adjacent to the line.  Note the
  difference in \ion{Mg}{2} flux between the two fits.\label{fig1}}  
\end{figure}

\clearpage 

\begin{figure}
\plotone{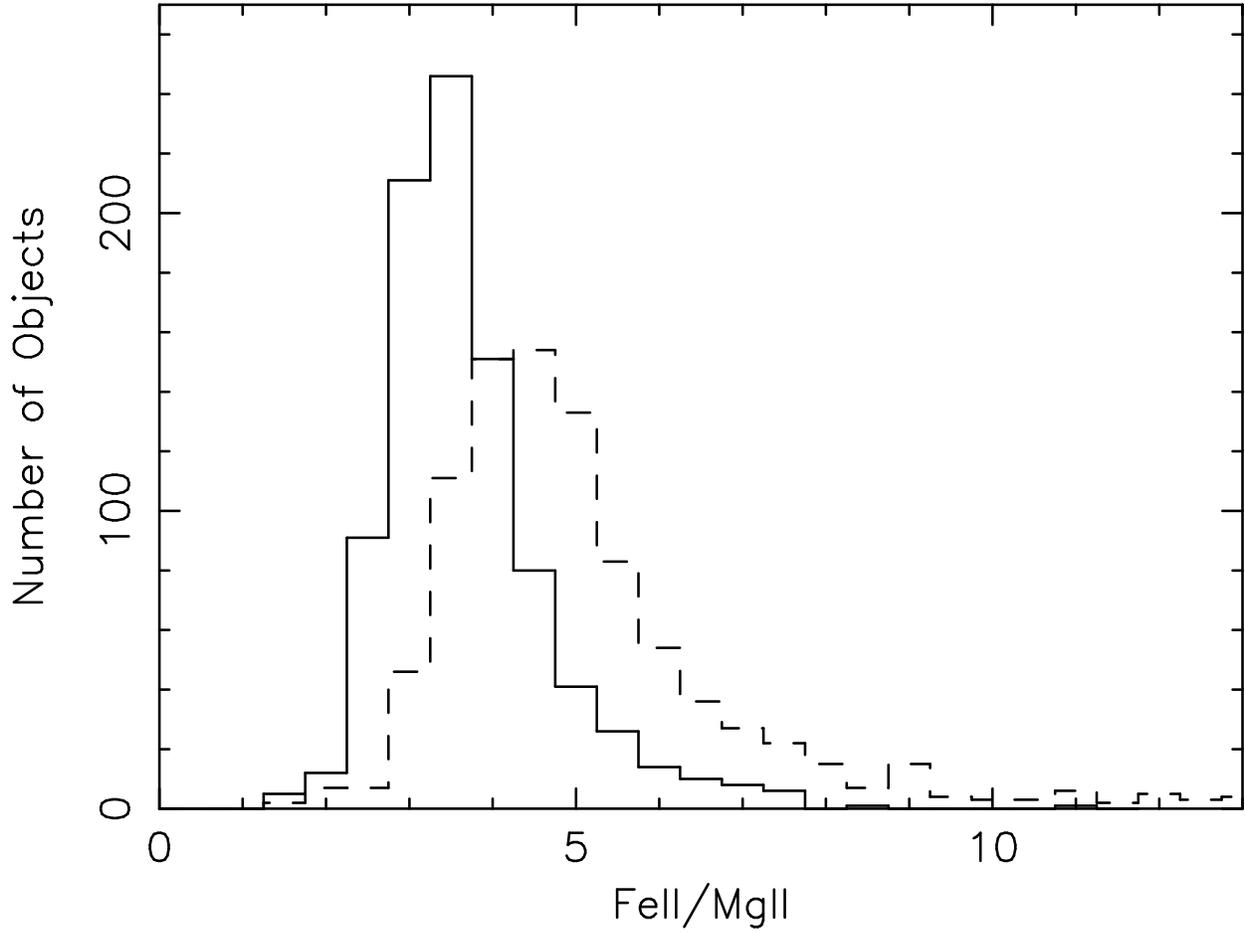}
\caption{Histograms of the measured \ion{Fe}{2}/\ion{Mg}{2}
  ratios. Solid: traditional \ion{Fe}{2} template; Dashed: new
  \ion{Fe}{2} template.  Note that an exceptional object  with a
measure  \ion{Fe}{2}/\ion{Mg}{2} ratio of 33.5 for the new template is
  not   shown.  \label{fig2}}  
\end{figure}

\clearpage 

\begin{figure}
\plotone{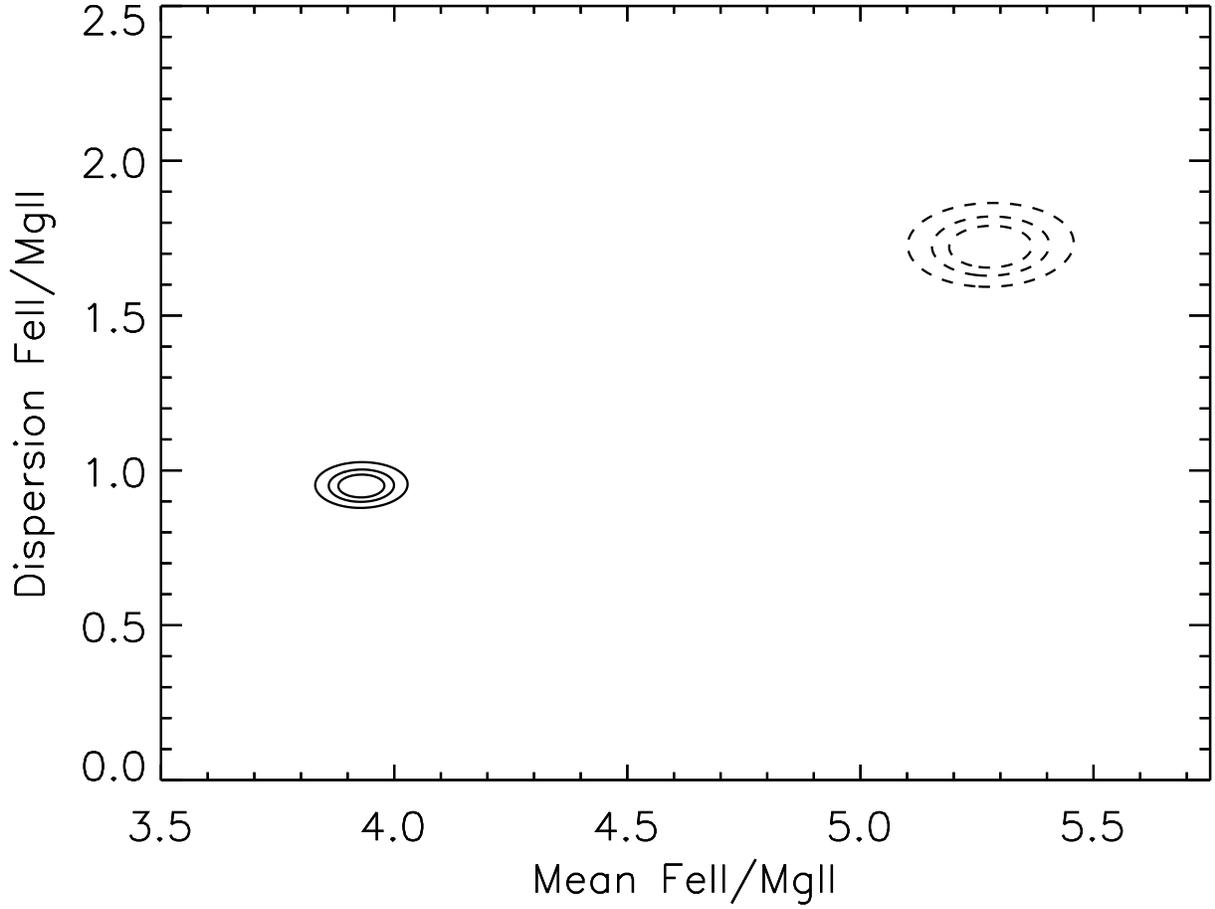}
\caption{Maximum likelihood-analysis contours for
\ion{Fe}{2}/\ion{Mg}{2}. Solid: traditional \ion{Fe}{2} template;
Dashed: new \ion{Fe}{2} template.  Contours are 68\%, 90\%, and 99\%
for two parameters of interest.  The fact that the dispersion in this
parameter is significantly greater than zero demonstrates that there
is a significant range in this ratio in the sample.\label{fig3}}
\end{figure}

\clearpage

\begin{figure}
\plotone{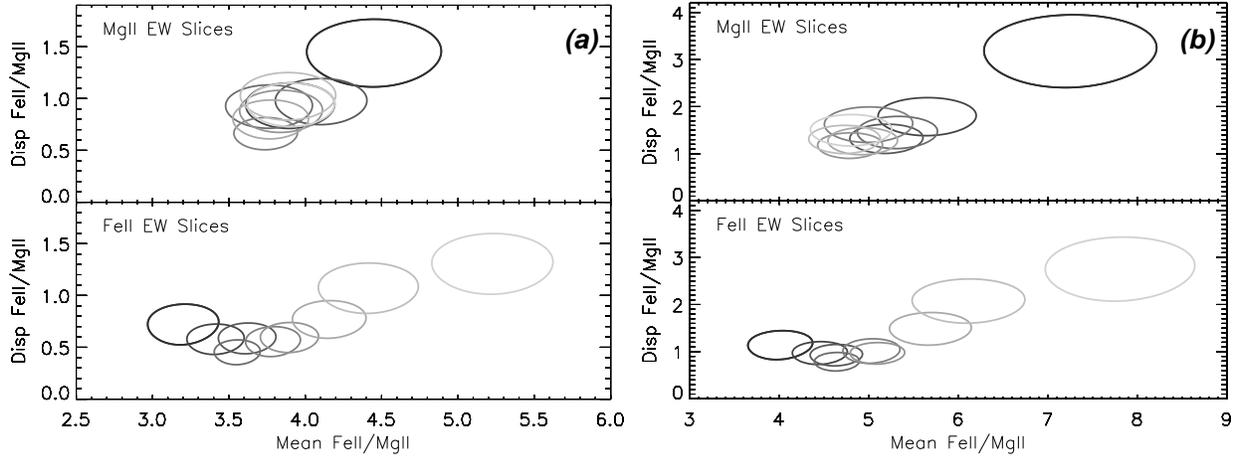}
\caption{Maximum likelihood contours of the mean and dispersion of
  \ion{Fe}{2}/\ion{Mg}{2} from nine bins from sorted \ion{Mg}{2}
  equivalent width (top) and sorted \ion{Fe}{2} equivalent width
  (bottom).  Only the 99\% contour is shown for clarity.  Contour
  shade gradient runs from dark to light for small to large values of
  the parameter. (a.) Results from fits using the
traditional \ion{Fe}{2} template; (b.) Results from fits using the new
\ion{Fe}{2} template. \label{fig4}} 
\end{figure}

\clearpage

\begin{figure}
\plotone{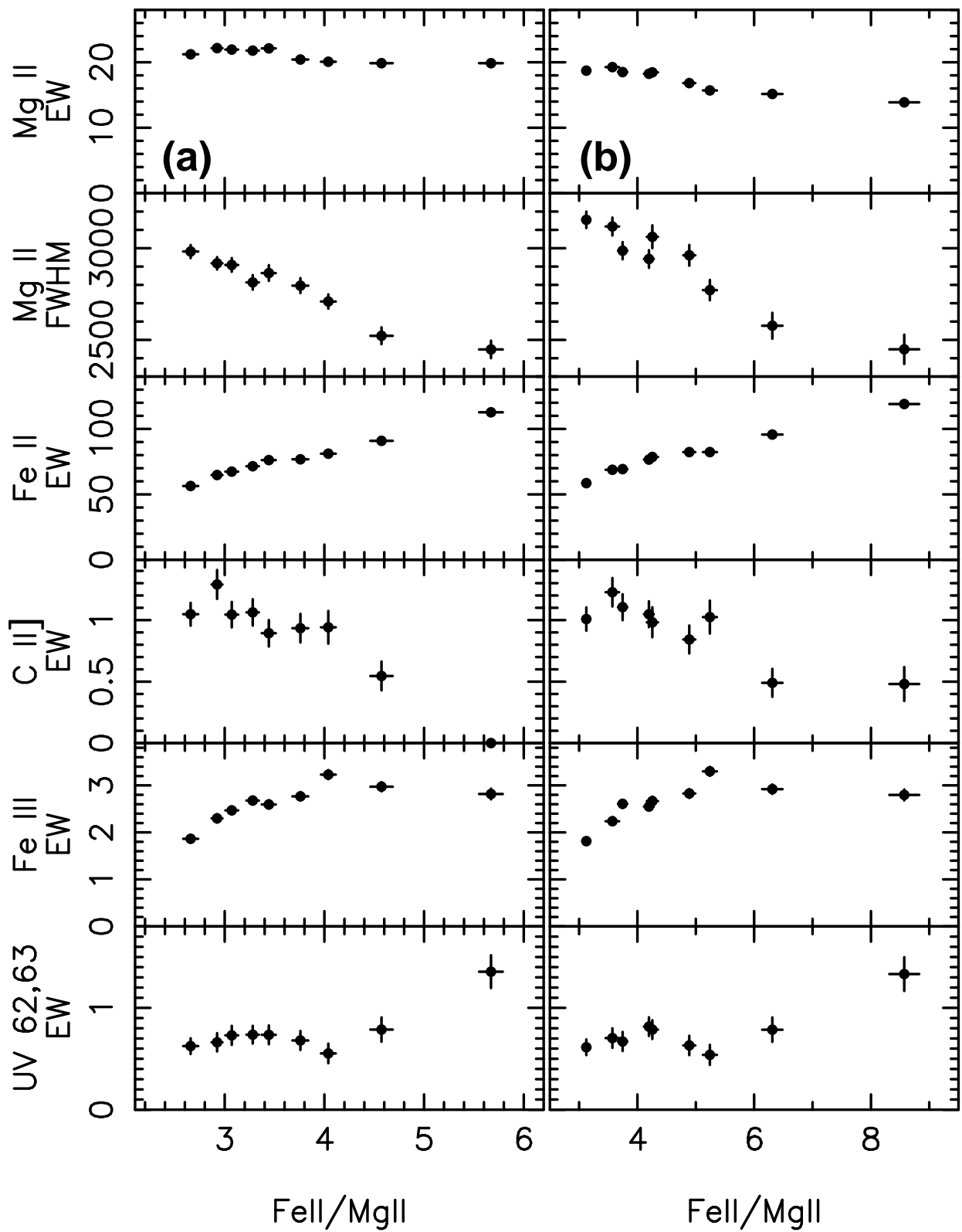}
\caption{Spectral fitting results for the nine composite spectra
obtained by  sorting \ion{Fe}{2}/\ion{Mg}{2} values, splitting into nine
bins, and accumulating spectra that have signal-to-noise ratio
greater than the median value.  (a.) Results from fits using the
traditional \ion{Fe}{2} template; (b.) Results from fits using the new
\ion{Fe}{2} template.\label{fig5}} 
\end{figure}

\clearpage 

\begin{figure}
\plotone{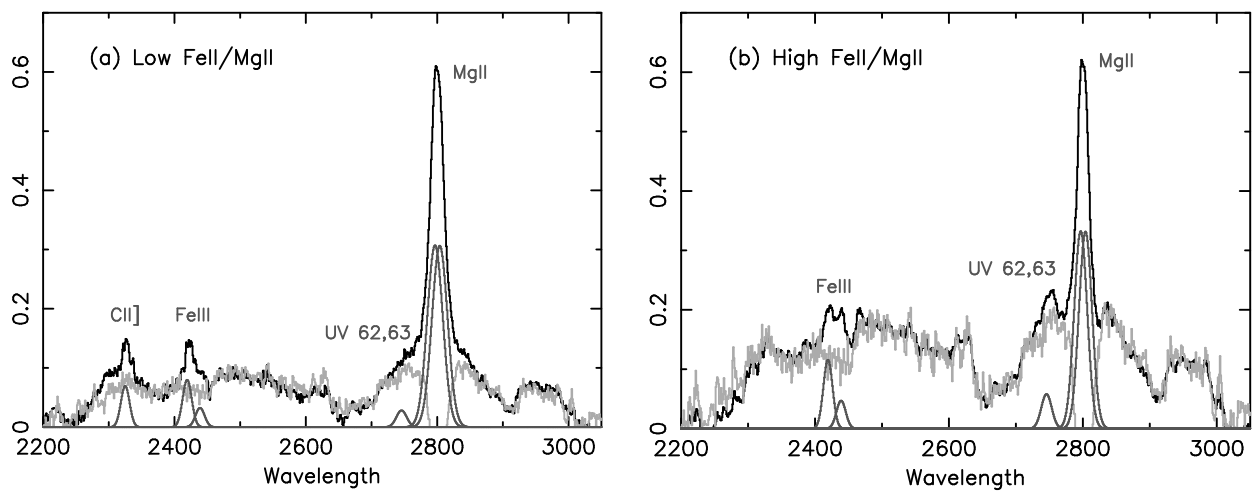}
\caption{Fits to the two extreme of nine composite spectra accumulated
  from sorted \ion{Fe}{2}/\ion{Mg}{2}. Only the results using the
  traditional template are shown; the results using the new template
  are essentially identical.  (a.) Composite
  spectrum from the bin containing the lowest values of
  \ion{Fe}{2}/\ion{Mg}{2}. (b.)  Composite
  spectrum from the bin containing the highest values of
  \ion{Fe}{2}/\ion{Mg}{2}.  \ion{C}{2}] is much stronger in the low
  \ion{Fe}{2}/\ion{Mg}{2} composite.  \label{fig6}} 
\end{figure}

\clearpage 

\begin{figure}
\plotone{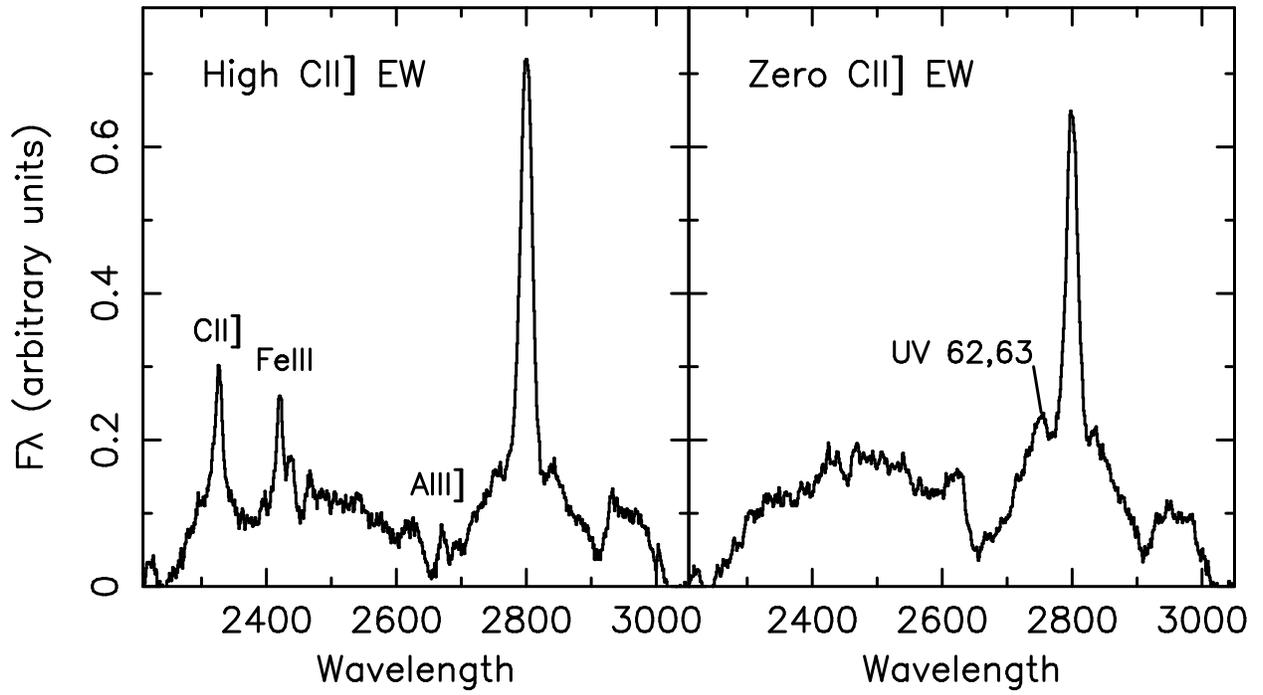}
\caption{Composite spectra compiled from spectra with high (left) and
  low (right) equivalent width \ion{C}{2}] lines. \label{fig7}} 
\end{figure}

\clearpage 

\begin{figure}
\plotone{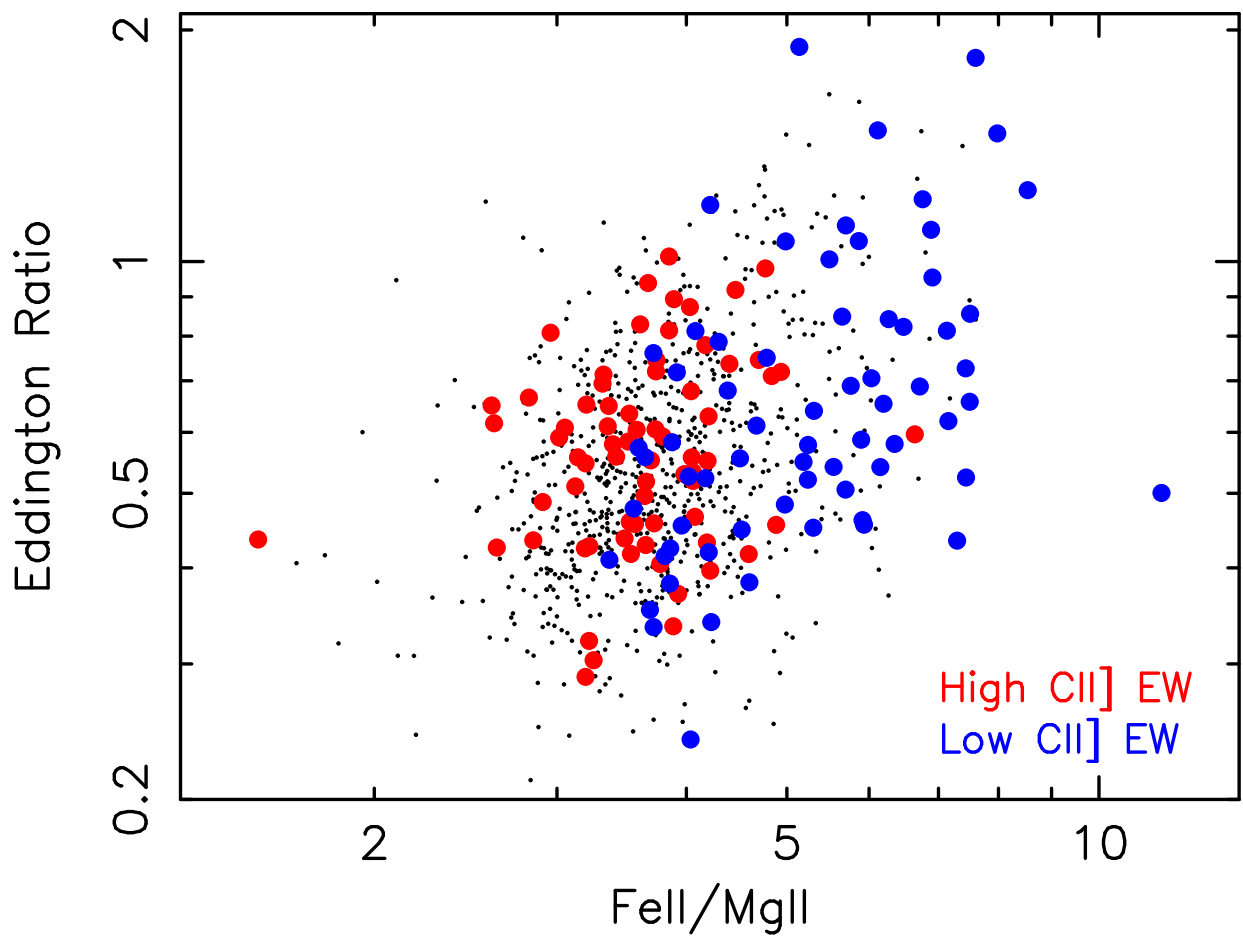}
\caption{Objects with zero \ion{C}{2}] equivalent width (large blue
  symbols) are more likely to have high \ion{Fe}{2}/\ion{Mg}{2}
  ratios, while objects with high \ion{C}{2}] equivalent widths (large
  red symbols) have low or average \ion{Fe}{2}/\ion{Mg}{2} ratios.
  Since \ion{C}{2}] is sensitive to density, this difference suggests
  that   the range in \ion{Fe}{2}/\ion{Mg}{2} ratios observed in the sample
  is a consequence of a range of physical conditions in the emitting
  gas.    Small symbols show the remaining objects in the sample that
  either have moderate \ion{C}{2}] equivalent width lines, or have
  zero \ion{C}{2}] equivalent width lines and low signal-to-noise
  ratio spectra. 
\label{fig8}}  
\end{figure}

\clearpage 

%%%UCP%%%

%% If you are not including electronic art with your submission, you may
%% mark up your captions using the \figcaption command. See the 
%% No more than seven \figcaption commands are allowed per page, 
%% so if you have more than seven captions, insert a \clearpage 

%% Tables should be submitted one per page, so put a \clearpage before

%% Two options are available to the author for producing tables:  the
%% deluxetable environment provided by the AASTeX package or the LaTeX
%% table environment.  Use of deluxetable is preferred.
%%

%% Three table samples follow, two marked up in the deluxetable environment,
%% one marked up as a LaTeX table.

%% In this first example, note that the \tabletypesize{}
%% command has been used to reduce the font size of the table.
%% Note also that the \label command needs to be placed 

\begin{deluxetable}{lllll}
\tabletypesize{\normalsize}

\tablewidth{0pc}
\tablenum{1}
\tablecaption{Maximum Likelihood Analysis}
\tablehead{ & \multicolumn{2}{c}{Traditional \ion{Fe}{2}
    Template} & \multicolumn{2}{c}{New \ion{Fe}{2} Template} \\
& \multicolumn{2}{c}{\hrulefill} & \multicolumn{2}{c}{\hrulefill} \\
 \colhead{Parameter} & \colhead{Mean} & \colhead{1$\sigma$ Dispersion} & \colhead{Mean} &
  \colhead{1$\sigma$ Dispersion}} 
\startdata
\ion{Mg}{2} EW (\AA\/) & $22.42^{+0.59}_{-0.61}$ &  $5.85^{+0.47}_{-0.41}$ & 
$18.00 \pm 0.50$  &   $4.88^{+0.39}_{-0.35}$ \\

$\log \rm L_{MgII}$ & $42.80 \pm 0.02$ &  $0.22 \pm 0.02$ &
$42.70 \pm 0.02$ &   $0.22 \pm 0.02$ \\

\ion{Fe}{2} EW (\AA\/) & $87.7 \pm {3.1}$  &  $30.4^{+2.5}_{-2.2}$ & $92.8 \pm
3.3$  &     $32.1^{+2.6}_{-2.3}$ \\

$\log \rm L_{FeII}$ &  $43.39 \pm 0.03$ &  $0.25 \pm 0.02$
& $43.41 \pm 0.03$ &  $0.25 \pm 0.02$ \\

\ion{Fe}{2}/\ion{Mg}{2} & $3.93 \pm 0.10$ &
$0.95^{+0.08}_{-0.07}$ & $5.28 \pm 0.18$  &
$1.72^{+0.14}_{-0.13}$ \\

\ion{Mg}{2}/\ion{Fe}{2} & $0.263 \pm 0.006$ &
$0.053^{+0.005}_{-0.004}$ & $0.202^{+0.006}_{-0.005}$ &  $0.051 \pm
0.004$ \\

\ion{Mg}{2} FWHM ($\rm\, km\, s^{-1}$) & $2635 \pm 35$ &
$340^{+30}_{-20}$ & $2576 \pm 48$   &    $460^{+40}_{-30}$ \\

$\log \rm L_{2500}$ & $41.53 \pm +0.02$ &  $0.21^{+0.02}_{-0.01}$
& $41.53^{+0.022}$  &    $0.21^{+0.02}_{-0.01}$ \\

$M_{BH}$ & $7.31 \pm 0.28 \times 10^7$ &  $2.73^{+0.21}_{-0.19}\times
10^7$ & $7.08^{+0.33}_{-0.32} \times 10^7$ & $3.17^{+0.25}_{-0.23}
\times 10^7$\\ 

Eddington Ratio &  $0.58 \pm 0.02$  &
$0.21 \pm 0.02$ &  $0.62\pm 0.03$   & $0.28 \pm 0.02$  \\
\enddata
\tablecomments{Quoted uncertainties are 99\% confidence for two
  parameters of interest ($\Delta\chi^2=9.21$).}
\end{deluxetable}

\clearpage

\begin{deluxetable}{lrrrrrrrrrrrr}
\tabletypesize{\scriptsize}
\rotate

\tablewidth{0pc}
\tablenum{2}
\tablecaption{Correlation Matrix}
\tablehead{ & \colhead{\ion{Mg}{2} EW} & \colhead{\ion{Fe}{2} EW} &
  \colhead{$\log(L_{MgII})$} & \colhead{$\log(L_{FeII})$} &
  \colhead{\ion{Fe}{2}/\ion{Mg}{2}} &
  \colhead{\ion{Mg}{2}/\ion{Fe}{2}} & 
  \colhead{\ion{Mg}{2} FWHM} & \colhead{$\rm L_{2500}$} &
  \colhead{$M_{BH}$} &   \colhead{$L/L_{Edd}$} }
\startdata
\ion{Mg}{2} EW & 1.00 &   0.69 &   0.28 &   0.20 &  $-$0.13 & 0.13 &
0.29 &  $-$0.23 &   0.05 &  $-$0.41 \\ 
& 1.00 &   0.51 &   0.30 &   0.09 &  $-$0.33 & 0.33 &   0.36 &  $-$0.24 &
0.16 &  $-$0.48 \\

\ion{Fe}{2} EW & 0.69 &   1.00 &   0.28 &   0.48 &   0.56 & $-$0.56 &
0.02 &  $-$0.08 &  $-$0.04 &  $-$0.09 \\
& 0.51 &   1.00 &   0.21 &   0.49 &   0.56 & $-$0.56 &  $-$0.01 &  $-$0.08 &
$-$0.06 &  $-$0.05 \\ 

$\log(L_{MgII})$ & 0.28 &   0.28 &   1.00 &   0.90 &   0.06 & $-$0.06 &
0.32 &   0.84 &   0.73 &   0.37 \\
& 0.30 &   0.21 &   1.00 &   0.83 &  $-$0.07 & 0.07 &   0.37 &   0.82 &
0.70 &   0.17 \\ 
 
$\log(L_{FeII})$ & 0.20 &   0.48 &   0.90 &   1.00 &   0.43 & $-$0.43 &
0.16 &   0.79 &   0.58 &   0.47 \\ 
& 0.09 &   0.49 &   0.83 &   1.00 &   0.43 & $-$0.43 &   0.14 &   0.79 &
0.49 &   0.37 \\

\ion{Fe}{2}/\ion{Mg}{2} & $-$0.13 &   0.56 &   0.06 &   0.43 &   1.00 &
$-$1.00 &  $-$0.32 &   0.12 &  $-$0.14 &   0.34 \\ 
& $-$0.33 &   0.56 &  $-$0.07 &   0.43 &   1.00 & $-$1.00 &  $-$0.37 &   0.12
&  $-$0.23 &   0.40 \\ 

\ion{Mg}{2}/\ion{Fe}{2} & 0.13 &  $-$0.56 &  $-$0.06 &  $-$0.43 &  $-$1.00 &
1.00 &   0.32 &  $-$0.12 &   0.14 &  $-$0.34 \\
& 0.33 &  $-$0.56 &   0.07 &  $-$0.43 &  $-$1.00 & 1.00 &   0.37 &  $-$0.12 &
0.23 &  $-$0.40 \\

\ion{Mg}{2} FWHM & 0.29 &   0.02 &   0.32 &   0.16 &  $-$0.32 & 0.32 &
1.00 &   0.18 &   0.78 &  $-$0.64 \\ 
& 0.36 &  $-$0.01 &   0.37 &   0.14 &  $-$0.37 & 0.37 &   1.00 &   0.17 &
0.86 &  $-$0.77 \\ 

$\rm L_{2500}$ & $-$0.23 &  $-$0.08 &   0.84 &   0.79 &   0.12 & $-$0.12 &
0.18 &   1.00 &   0.71 &   0.59 \\
& $-$0.24 &  $-$0.08 &   0.82 &   0.79 &   0.12 & $-$0.12 &   0.17 &   1.00
&   0.62 &   0.44 \\ 

$M_{BH}$ & 0.05 &  $-$0.04 &   0.73 &   0.58 &  $-$0.14 & 0.14 &   0.78 &
0.71 &   1.00 &  $-$0.07  \\
& 0.16 &  $-$0.06 &   0.70 &   0.49 &  $-$0.23 & 0.23 &   0.86 &   0.62 &
1.00 &  $-$0.36 \\ 

$\rm L/L_{Edd}$ & $-$0.41 & $-$0.09 & 0.37 & 0.47 & 0.34 & $-$0.34 & $-$0.64 &
0.59 & $-$0.07 & 1.00 \\
& $-$0.48 & $-$0.05 & 0.17 & 0.37 & 0.40 & $-$0.40 & $-$0.77 & 0.44 & $-$0.36 &
1.00 \\

\enddata
\tablecomments{Correlation coefficient is Spearman Rank. For
  each entry, the upper and lower numbers are from analyses using the
  traditional and new iron templates, respectively.}
\end{deluxetable}

\clearpage

\begin{deluxetable}{lrrrrr}
\tabletypesize{\scriptsize}
\rotate

\tablewidth{0pc}
\tablenum{3}
\tablecaption{Simulation Results}
\tablehead{\colhead{Parameter\tablenotemark{a}} & \colhead{Real} & 
\colhead{Simulation 1\tablenotemark{b}} & 
\colhead{Simulation 2\tablenotemark{c}} & 
\colhead{Simulation 3\tablenotemark{d}} & 
\colhead{Simulation 4\tablenotemark{e}}}
\startdata

\cutinhead{Distribution of Data} \\

\ion{Mg}{2} EW Mean$\pm$ Standard Dev. & $22.5 \pm 6.1$
&  $22.6 \pm 6.4$ & $22.7 \pm 7.0$  & $22.4 \pm 5.5$ & $22.7 \pm 7.0$ \\

ML \ion{Mg}{2} EW Mean$\pm$ Standard Dev. & $22.4 \pm
5.9$ & $22.5 
\pm 6.3$ & $22.6 \pm 6.8$ & $22.2 \pm 5.2$ & $22.5 \pm 6.5$ \\

\ion{Fe}{2} EW Mean$\pm$ Standard Dev. & $88 \pm 32$ &  $88 \pm 30$ &
$87 \pm 28$ & $87 \pm 27$ & $87 \pm 27$ \\ 

ML \ion{Fe}{2} EW Mean$\pm$ Standard Dev. & $88 \pm 30$ & $87 \pm 29$
& $87 \pm 27$ & $87 \pm 27$ & $86 \pm 26$ \\

\ion{Fe}{2}/\ion{Mg}{2} Mean$\pm$ Standard Dev. & $3.9 \pm 1.0$ & $3.9
\pm 1.0$ & $3.95 \pm 0.96$ & $3.95 \pm 0.95$ & $3.95 \pm 0.96$ \\

ML \ion{Fe}{2}/\ion{Mg}{2} Mean$\pm$ Standard Dev. & $3.93 \pm 0.95$ &
$3.92 \pm 0.91$ & $3.92 \pm 0.91$ & $3.90 \pm 0.89$ & $3.90 \pm 0.89$
\\ 

\cutinhead{Derived Distribution Properties} \\

\ion{Mg}{2} EW Standard Dev./Mean & 0.27 & 0.28 & 0.31 & 0.25 & 0.31 \\

ML \ion{Mg}{2} EW Standard Dev./Mean & 0.26 & 0.28 & 0.30 & 0.23 &
0.29 \\

\ion{Fe}{2} EW Standard Dev./Mean & 0.36 & 0.34 & 0.32 & 0.31 & 0.31
\\ 

ML \ion{Fe}{2} EW Standard Dev./Mean & 0.35 & 0.33 & 0.31 & 0.31 &
0.30 \\ 

\ion{Fe}{2}/\ion{Mg}{2} Standard Dev./Mean & 0.26 & 0.26 & 0.24 & 0.24
& 0.24 \\ 

ML \ion{Fe}{2}/\ion{Mg}{2} Standard Dev./Mean & 0.24 & 0.23 & 0.23 &
0.23  & 0.23 \\

Mean \ion{Mg}{2} EW Error/EW & 0.040 & 0.037 & 0.036 & 0.059 & 
0.059 \\

Mean \ion{Fe}{2} EW Error/EW & 0.064 & 0.060 & 0.059 & 0.059 & 0.060 \\

\cutinhead{Correlations} \\

Correlation between \ion{Fe}{2} EW and \ion{Fe}{2}/\ion{Mg}{2} & 0.56
& 0.57 & 0.43 & 0.61 & 0.42 \\

Correlation between \ion{Mg}{2} EW and \ion{Mg}{2}/\ion{Fe}{2} & 0.13
& 0.21 & 0.36 & 0.24 & 0.41 \\

\enddata
\tablenotetext{a}{ML stands for maximum likelihood.  Means and
  standard deviations are estimated two ways: first, as the straight
  mean and standard deviation, and second, using the maximum
  likelihood technique.}
\tablenotetext{b}{Simulations designed to have same equivalent width
  distributions and same mean relative uncertainty as the real data.}
\tablenotetext{c}{Simulations in which the \ion{Mg}{2} equivalent
  width distribution is the same as the \ion{Fe}{2} equivalent width
  distribution (see maximum likelihood mean/standard dev.), but same
  relative uncertainty as the real data.}
\tablenotetext{d}{Simulations in which the equivalent width
  distributions are the same as the real data, but in which the
  uncertainty in the \ion{Mg}{2} equivalent width has been increased
  so that the relative uncertainty is the same as that of
  \ion{Fe}{2} (see mean EW error/EW). } 
\tablenotetext{e}{Simulations in which the \ion{Mg}{2} equivalent
  width distribution is the same as the \ion{Fe}{2} equivalent width
  distribution (see  mean/standard dev.), and the
  uncertainty in the \ion{Mg}{2} equivalent width has been increased
  so that the relative uncertainty is the same as that of
  \ion{Fe}{2} (see mean EW error/EW).}

\end{deluxetable}

\end{document}